\documentclass[%
 reprint,
 superscriptaddress,
 showpacs,preprintnumbers,
 amsmath,amssymb,
 aps,
 prl,
]{revtex4-1}

\usepackage{graphicx}
\usepackage{dcolumn}
\usepackage{bm}
\usepackage{makecell}
\usepackage{braket} 
\usepackage{siunitx}
\usepackage{color}
\usepackage{xr}
\usepackage{upgreek}
\usepackage{hyperref}   

\begin{document}

\title{Characterizing and optimizing qubit coherence based on SQUID geometry}

\def\RLEaffil{Research Laboratory of Electronics, Massachusetts Institute of Technology, Cambridge, MA 02139, USA}
\def\LLaffil{MIT Lincoln Laboratory, Lexington, MA 02421, USA}
\def\Physaffil{Department of Physics, Massachusetts Institute of Technology, Cambridge, MA 02139, USA}
\def\EECSaffil{Department of Electrical Engineering and Computer Science, Massachusetts Institute of Technology, Cambridge, MA 02139, USA}

\author{Jochen Braum\"uller$^{\dagger}$}
\thanks{These two authors contributed equally.\\$^{\dagger}$ jbraum@mit.edu}
\author{Leon Ding}
\thanks{These two authors contributed equally.\\$^{\dagger}$ jbraum@mit.edu}
\author{Antti Veps\"al\"ainen}
\author{Youngkyu Sung}
\author{Morten Kjaergaard}
\affiliation{\RLEaffil}
\author{Tim Menke}
\affiliation{\RLEaffil}
\affiliation{\Physaffil}
\author{Roni Winik}
\affiliation{\RLEaffil}
\author{David Kim}
\author{Bethany M. Niedzielski}
\author{Alexander Melville}
\author{Jonilyn L. Yoder}
\author{Cyrus F. Hirjibehedin}
\affiliation{\LLaffil}
\author{Terry P. Orlando}
\author{Simon Gustavsson}
\affiliation{\RLEaffil}
\author{William D. Oliver}
\affiliation{\RLEaffil}
\affiliation{\Physaffil}
\affiliation{\LLaffil}
\affiliation{\EECSaffil}

\date{\today}

\newcommand*{\I}{\mathrm{i}}
\newcommand*{\totd}{\text{d}}

\newcommand{\RN}[1]{\textup{\uppercase\expandafter{\romannumeral#1}}}

\begin{abstract}

The dominant source of decoherence in contemporary frequency-tunable superconducting qubits is 1/$f$ flux noise. To understand its origin and find ways to minimize its impact, we systematically study flux noise amplitudes in more than 50 flux qubits with varied SQUID geometry parameters and compare our results to a microscopic model of magnetic spin defects located at the interfaces surrounding the SQUID loops. Our data are in agreement with an extension of the previously proposed model, based on numerical simulations of the current distribution in the investigated SQUIDs. Our results and detailed model provide a guide for minimizing the flux noise susceptibility in future circuits.

\end{abstract}

\maketitle

Superconducting circuits are leading candidates to implement quantum hardware capable of performing certain computational tasks more efficiently than classical computers~\cite{Feynman1982,Lloyd1996}. During the last two decades -- and lately at a more rapid pace -- quantum circuits have become increasingly complex~\cite{Devoret2013,Kjaergaard2019}. This has enabled several proof-of-principle demonstrations of small quantum algorithms and simulations, heralding the era of noisy intermediate scale quantum (NISQ) devices~\cite{Kjaergaard2019} and recently, a demonstration of quantum advantage in sampling the output distribution of a pseudo-random quantum circuit~\cite{Boixo2018,Arute2019}. However, a major roadblock toward scaling superconducting circuits to perform useful computations is the limited qubit coherence~\cite{Oliver2013}, restricting run times of algorithms or simulations and creating a large resource overhead in quantum error correction schemes.

With many of the recently implemented circuits relying on frequency-tunable qubits, the dominant source of dephasing in these qubits~\cite{Barends2013, Arute2019} is low-frequency flux noise with a power spectral density (PSD) that is inversely proportional to frequency~\cite{Paladino2014}. Such $1/f$ noise is ubiquitous in condensed matter systems~\cite{Dutta1981} and was observed in the context of Josephson devices more than three decades ago~\cite{Koch1983}. With the advent of superconducting qubits, $1/f$ noise in superconducting quantum interference devices (SQUIDs) has been shown to cause qubit dephasing ~\cite{Yoshihara2006,Kakuyanagi2007,Koch2007b,Bialczak2007,Gustavsson2011,Yan2013,Krantz2019} as well as qubit relaxation~\cite{Yan2016, Quintana2017}. It was proposed that $1/f$ flux noise in qubits comprising SQUIDs originates from magnetic two-level system defects residing in the oxide layers surrounding the SQUID loops~\cite{Koch2007b}. The model assumes a temperature-activated flipping of independent electronic spins that are randomly oriented and have a random energy distribution~\cite{Dutta1981}, leading to a $1/f$ noise PSD. These spin entities can either be single electrons or spin clusters which form a collective spin. Oxygen adsorbates were determined to be candidate sources for such spin defects by density functional theory calculations~\cite{Wang2015} and x-ray spectroscopy~\cite{Kumar2016}.

An analytic approximation of this microscopic model has been derived by Bialczak \textit{et al.}~\cite{Bialczak2007}, yielding an expression for the noise PSD $S(\omega)\propto R/W$, where $R$ is the radius of the SQUID loop and $W$ is the width of the superconducting strip forming the SQUID. While indications of a correct scaling with wire width have been reported~\cite{Lanting2009}, experiments with superconducting flux qubits or phase qubits could not quantitatively confirm the noise amplitudes predicted by the model~\cite{Lanting2009,Bialczak2007}, and the formation of spin clusters was proposed in order to reconcile the observed noise levels with the model~\cite{Sendelbach2008,Anton2013}. The origin of $1/f$ noise in SQUIDs has remained an unsolved question.

\begin{figure*}
\includegraphics{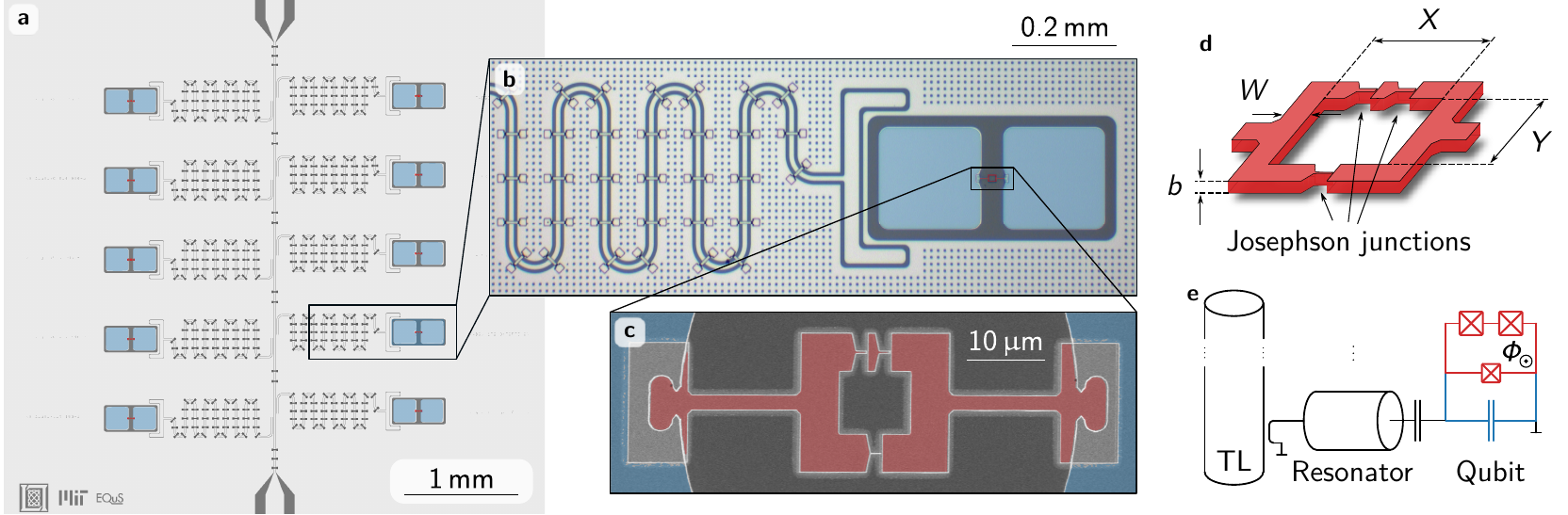}
\caption{Noise spectroscopy device. (a) Each chip holds ten uncoupled capacitively shunted flux qubits with individual readout resonators, featuring five different SQUID loop variations at a two-fold redundancy. (b) Optical micrograph of one of the qubits and part of its readout resonator. The capacitive shunt is colored in blue. (c) Electron microscope image of a fabricated SQUID loop. (d) Schematic representation of the SQUID parameters varied across different designs. The SQUID dimensions $X$, $Y$ are measured along the inner edge of the SQUID, $W$ is the width of the superconducting leads, and $b$ is the film thickness. (e) Effective schematic for one qubit-resonator pair coupled to the common transmission line (TL).}
\label{fig:device}
\end{figure*}

In this Letter, we study $1/f$ flux noise in more than 50 capacitively shunted flux qubits~\cite{Yan2016} with systematically varied geometric parameters of their SQUID loops. Our data show quantitative agreement with the proposed microscopic model of independent magnetic defects that reside in the interface layers surrounding the SQUIDs; in particular we demonstrate that the extracted flux noise amplitudes follow the expected trends over a large SQUID parameter regime. Since the analytic approximation~\cite{Bialczak2007} of the model is of limited applicability and accuracy for realistic circuit geometries, we present a numerical extension to the model, taking into account details of the geometry of generalized SQUID loops.

Our experiment incorporates results from six different samples comprising ten uncoupled capacitively shunted flux qubits~\cite{Yan2016} each, see Fig.~\ref{fig:device}. Qubit control and dispersive state readout is performed through individual capacitively coupled $\lambda/4$ waveguide resonators, which are in turn inductively coupled to a common $\SI{50}{\ohm}$ transmission line. The samples are cooled down to approximately $\SI{10}{mK}$ in a dilution refrigerator. Microwave transmission through the transmission line is used to projectively measure the qubit state with a heterodyne detection scheme at room temperature. Details on sample fabrication are provided in Supplementary Sec.~\RN{1}.

With the Hamiltonian parameters of each flux qubit nominally identical, we vary the geometric parameters of their SQUIDs as illustrated in Fig.~\ref{fig:device}(d). While the thickness $b=\SI{190}{nm}$ of the bilayer aluminum film is fixed, the side lengths $X$ and $Y$, referenced to the inner circumference, and the wire width $W$ are varied. Every sample varies either the inner SQUID perimeter $2X + 2Y$, the aspect ratio $X/Y$, or the width $W$. In order to reduce systematic errors, each SQUID variant is represented twice within a chip, resulting in five distinct SQUID geometries per chip. The ranges of the parameter variations are centered around state-of-the-art values used in high-coherence flux qubits~\cite{Yan2016}, $X=\SI{9}{\micro m}$, $Y=\SI{8}{\micro m}$, and $W=\SI{1}{\micro m}$. Figure~\ref{fig:device}(e) shows the effective circuit schematic for one qubit-resonator pair coupled to the common transmission line. Circuit parameters are summarized in Supplementary Sec.~\RN{2}. A global external flux bias $\Phi$ is applied to the SQUID loops with a coil located in the lid of the sample package.

\begin{figure} 
\includegraphics{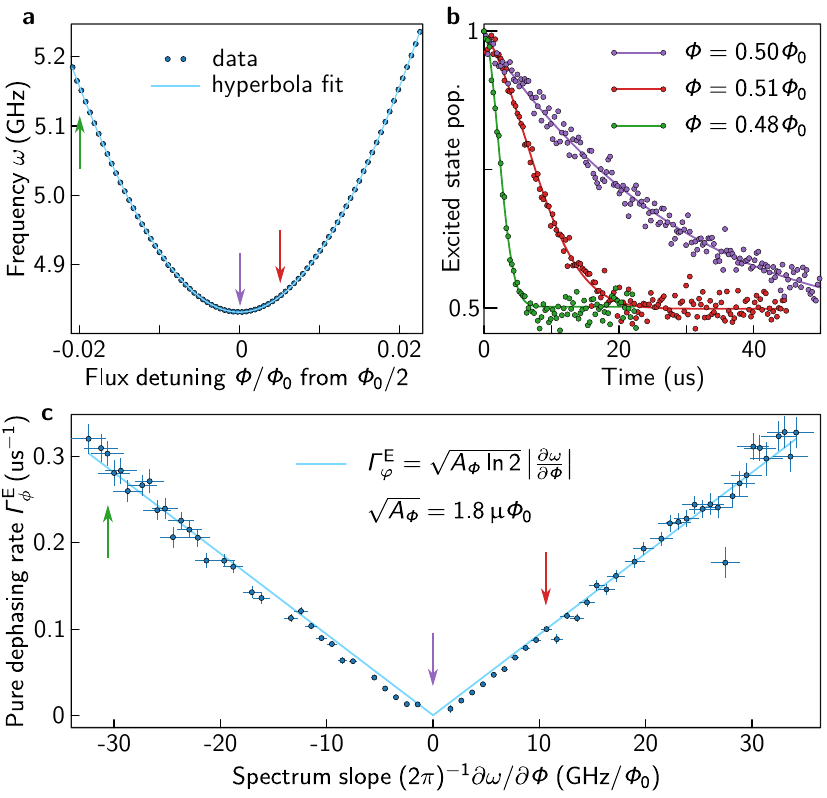}
\caption{Experimental technique used for noise spectroscopy. (a) Qubit spectrum around its flux sweet spot at $\Phi=\Phi_0/2$. A hyperbolic fit enables us to numerically extract the slope $(2\pi)^{-1}\partial\omega/\partial\Phi$ of the spectrum. (b) Spin echo dephasing traces at three illustrative locations of the spectrum (indicated by the arrows). (c) By plotting the extracted pure dephasing rates $\Gamma_\phi ^{\mathrm{E}}$ as a function of the spectrum slope, we can extract the $1/f$ flux noise amplitude $\sqrt{A_\Phi}$ from a linear fit.}
\label{fig:technique}
\end{figure}

We perform noise spectroscopy for every qubit using a sequence of measurements first demonstrated in Ref.~\cite{Yoshihara2006}. We first extract the qubit spectrum around the optimal bias point at $\Phi=\Phi_0/2$, see Fig.~\ref{fig:technique}(a). Subsequently, we measure qubit relaxation by exciting the qubit with a calibrated $\pi$-pulse and recording the residual excited state population after varying times. Finally, we perform a spin-echo experiment, where a $\pi$-pulse in the middle of a Ramsey sequence inverts the sign of the phase accrual rate due to quasi-static low-frequency noise. As shown in Fig.~\ref{fig:technique}(b), we observe an exponential decay function at the sweet spot, where decoherence is relaxation limited. Further away from the sweet spot, the decay function is predominantly Gaussian, indicative of pure dephasing due to $1/f$ noise~\cite{Makhlin2004,Yoshihara2006}. As detailed in Supplementary Sec.~\RN{3}, the Gaussian pure dephasing rate takes the form $\Gamma_\phi^{\mathrm{E}}=\sqrt{A_\Phi \ln2} \left|\partial\omega/\partial\Phi\right|$ for the echo experiment, assuming Gaussian statistics of the qubit phase accumulation~\cite{Makhlin2004} and a noise PSD $S_\Phi(\omega)=A_\Phi/|\omega|$ with noise amplitude $\sqrt{A_\Phi}$ at $\omega/2\pi=\SI{1}{Hz}$. To find the exponential decay rate $\Gamma_\mathrm{exp}$ and Gaussian dephasing rate $\Gamma_\phi^{\mathrm{E}}$, we perform a fit to the decay function $p(t) \propto \exp[-\Gamma_{\mathrm{exp}}t-(\Gamma_\phi^{\mathrm{E}}t)^2]$~\cite{Yoshihara2006}, where $\Gamma_\mathrm{exp}$ is kept a free parameter with an initial guess of $(2T_1)^{-1}$, as extracted in the preceding relaxation measurement. 
In order to numerically extract the slope of the spectrum, we fit the hyperbola $\hbar\omega(\Phi)=\sqrt{\Delta^2+\epsilon(\Phi)^2}$ to the data in Fig.~\ref{fig:technique}(a), which is a good approximation over the measured range close to the sweet spot~\cite{Yan2016}.

The pure dephasing rate $\Gamma_\phi^{\mathrm{E}}$ as a function of the slope of the spectrum $(2\pi)^{-1}\partial\omega/\partial\Phi$ for one of the measured qubits is shown in Fig.~\ref{fig:technique}(c). We perform two separate linear fits (for positive and negative slope) and extract the noise amplitude $\sqrt{A_\Phi}$ and its uncertainty from an error-weighted average. Since pure dephasing in the Gaussian approximation vanishes at the sweet spot we enforce an intercept with the origin. About $20\%$ of the qubits show a bending of data points to a finite (positive) dephasing rate near the sweet spot. We attribute these deviations to other higher-frequency dephasing processes, which do not significantly compromise the extracted noise amplitude. It is important to note that the validity of our experimental procedure is limited to a noise PSD $S(\omega)\propto \omega^{-\alpha}$ with $\alpha=1$, see Supplementary Sec.~\RN{3}. While $1/f$ noise has been observed with a scaling where $\alpha \leq 1$~\cite{Bylander2011,Ithier2005,Dutta1981,Paladino2014}, this assumption is compatible with previous experiments extrapolated to $\sim\SI{10}{mK}$~\cite{Anton2013} and is supported by the Gaussian decay function we observe in our experiment.

\begin{figure} 
\includegraphics{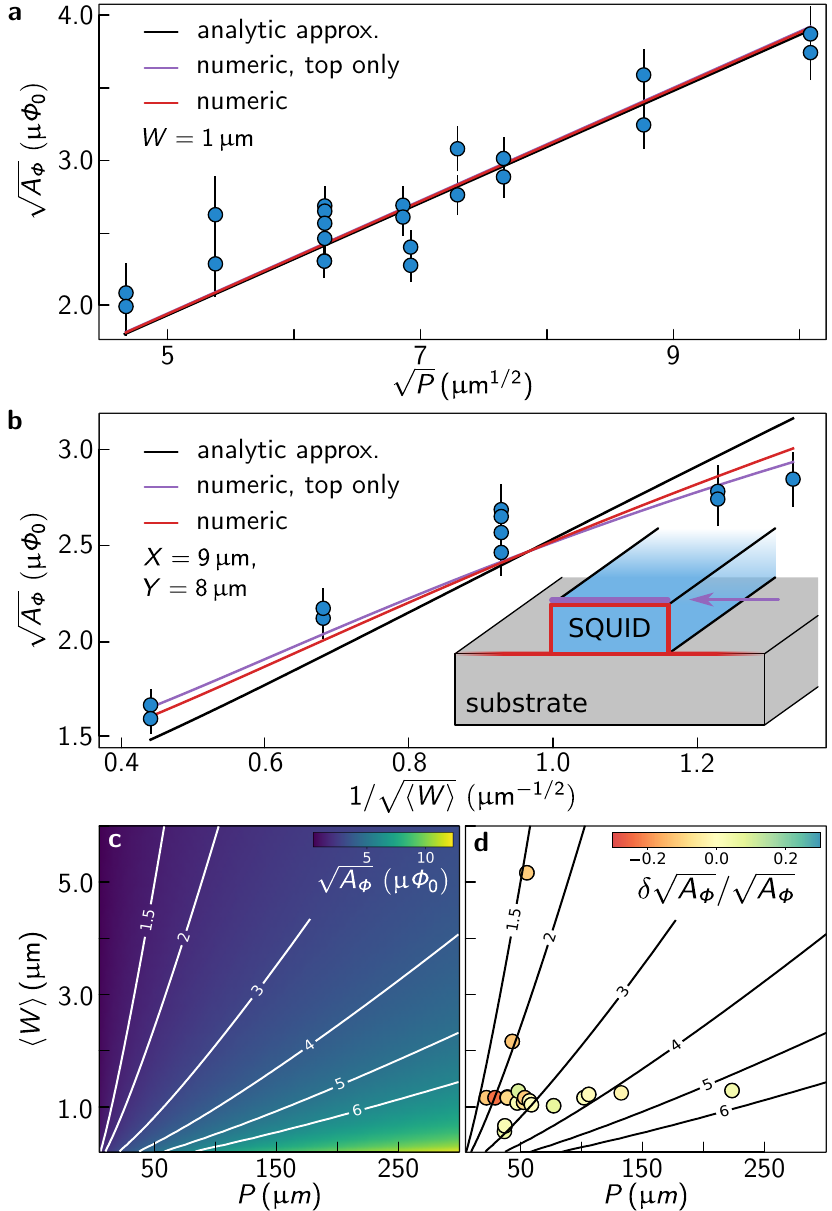}
\caption{Flux noise amplitudes $\sqrt{A_\Phi}$ as a function of SQUID geometry parameters for (a) constant wire width $W$ and (b) constant inner perimeter $2X+2Y$. Each line corresponds to an independent fit in this two-dimensional parameter space to the analytic approximation (black) and numeric variations (purple and red) of the model (see inset). (c) $\sqrt{A_\Phi}$ as a function of the effective width $\langle W\rangle $ and perimeter $P$ based on a fit to the numeric model taking into account all relevant interfaces (red lines in panels a and b). (d) Data points show experimentally investigated parameter combinations with relative deviations from the same numerical model color-coded.}
\label{fig:results}
\end{figure}

Figure~\ref{fig:results} shows the measured flux noise amplitudes $\sqrt{A_\Phi}$ as a function of SQUID geometry. We categorize the design variations into two groups. Qubits in the first group have SQUID loops with a constant wire width $W = \SI{1}{\micro m}$ but varying perimeters $\SI{21}{\micro m} \le P \le \SI{101}{\micro m}$, see Fig.~\ref{fig:results}(a). We define the perimeter $P=2X+2Y+4W$, measured along the center-line of the SQUID. The second group of measured qubits have SQUID loops with a fixed inner perimeter $2X+2Y=\SI{34}{\micro m}$ and varying wire width $\SI{0.4}{\micro m} \le W \le \SI{5}{\micro m}$, see Fig.~\ref{fig:results}(b). These sub-categories can be understood as line-cuts in the two dimensional parameter space $\sqrt{A_\Phi(P,W)}$, given in Fig.~\ref{fig:results}(c). 

Measured data show an approximately linear dependence of the noise power $A_\Phi$ on SQUID perimeter $P$ (Fig.~\ref{fig:results}(a)) and on the inverse wire width $W$ (Fig.~\ref{fig:results}(b)). By investigating SQUID loops of varying aspect ratio $X/Y$, we are able to confirm the linear scaling of $A_\Phi$ with SQUID perimeter rather than its area. Flux noise that is caused by fluctuations in the bias current source scales the noise amplitude as $\sqrt{A_\Phi}\propto\sqrt{\langle\Phi^2\rangle}\propto B \cdot A\propto P^2$, where $B$ is the induced magnetic field in the SQUID and $A$ its area. Since this is a different scaling than experimentally observed, we conclude that noise from the current source is insignificant for our experiment.

We compare our experimental data with a model that assumes $1/f$ flux noise to originate from local magnetic two-level system defects residing in the interface layers surrounding the qubit SQUID loops. This model had been proposed previously~\cite{Koch2007b,Bialczak2007} but has eluded quantitative experimental verification. The model assumes non-interacting magnetic defects of areal density $\sigma$ and with an average magnetic moment $m$, undergoing a thermally activated, uncorrelated flipping of their spin direction and thereby creating flux noise in the SQUID loop, ultimately leading to qubit decoherence. Modified to the rectangular geometry of the SQUIDs used in our experiment, see Fig.~\ref{fig:device}, the total flux variance $\langle \Phi^2 \rangle$ in the SQUID is
\begin{equation}
\langle \Phi^2 \rangle = \frac{\mu_0^2}{3 \pi} m^2 \sigma \frac{P}{W} \left( \frac{\ln(2 b W / \lambda^2)}{2 \pi} + \frac{e - 1}{2 \pi} \right),
\label{eq:model}
\end{equation}
where $\mu_0$ is the magnetic constant and $\lambda$ is the superconducting penetration depth of aluminum. The term in brackets is a result of assuming a surface current density $K(x)\propto 1/\sqrt{1-(2x/W)^2}$~\cite{Rhoderick1962} for $-W/2<x<W/2$, valid in the regime where the film thickness $b\sim\lambda$ and width $W\gg\lambda$. A detailed derivation can be found in Supplementary Sec.~\RN{4}A.

The scaling of flux noise with loop perimeter $P$ can be intuitively understood, since the total number of magnetic defects increases proportionally. The inverse scaling with wire width $W$ is less intuitive, given the increased number of participating defects for wider wires. It can be motivated by the following picture: for a constant persistent current in the SQUID, the magnetic field is diluted across more defect spins residing in the interfaces when increasing the wire width. Since the defects are uncorrelated, their contribution to the total flux noise partially cancels, resulting in an effective decrease of the total flux noise. Independent of the geometry, a uniform current density across the width of the SQUID arms minimizes the flux noise amplitude~\cite{optcurrent}. This provides an explanation for the previous observation that the presence of a superconducting ground plane reduces flux noise~\cite{Lanting2009,VanDuzer1981}.

In order to connect the noise amplitude $\sqrt{A_\Phi}$ extracted from measured data with the above model, we use $\langle\Phi^2\rangle = \int_{-\infty}^{\infty}\totd\omega S_\Phi(\omega) g_{\mathrm{E}}(\omega)=2A_\Phi\ln2$. To account only for frequencies our echo experiment is sensitive to, the integration is weighted by its filter function $g_{\mathrm{E}}$, see Supplementary Sec.~\RN{3}.

Both black lines in Fig.~\ref{fig:results}(a,b) belong to the same two-dimensional fit to the analytical approximation of the model in Eq.~\eqref{eq:model}, using only a single fit parameter $m^2 \sigma$. Assuming a penetration depth of $\lambda=\SI{40}{nm}$~\cite{Poole2013} and that the magnetic moment corresponds to a Bohr magneton, $m=\mu_{\mathrm{B}}$, we recover a surface spin density $\sigma= \SI{1.2e17}{m^{-2}}$, a factor of four off the previously predicted~\cite{Koch2007b} and observed~\cite{Sendelbach2008} value of $\SI{5e17}{m^{-2}}$. With an effective spin magnetic moment of $1.8\mu_{\mathrm{B}}$, as suggested for defects formed by oxygen adsorbates on the SQUID surface~\cite{Wang2015}, we extract $\sigma=\SI{3.7e16}{m^{-2}}$. 

Due to an offset between the bottom and top metallizations in the shadow evaporation process, the width along the vertical arms of the SQUIDs is increased. This effect is most pronounced in SQUIDs with small aspect ratios ($X\ll Y$), and it also has a noticeable effect on SQUIDs with thin wires. In order to account for this changing width in the SQUID loops, we plot an average width $\langle W\rangle$ in Fig.~\ref{fig:results}(b-d).

The analytic approximation of the model Eq.~\eqref{eq:model} is only valid in the regime where $b\sim\lambda$ and $W\gg\lambda$, but in our experiment, $b/\lambda\approx 5$. We attribute the deviations of data points in Fig.~\ref{fig:results}(b) from the linear scaling for thin wires (largest $1/\langle W\rangle$) to a partial breakdown of the approximate variant of the model. We extend the model by numerically computing the volume current density in the investigated SQUIDs, thereby overcoming the limitations of the analytic approximation. In our numerical approach, the arms of the SQUIDs are modeled as long superconducting strips, which are discretized into parallel sections. The currents in each segment are calculated based on the two-fluid model of superconductivity, where the supercurrent contribution is described through London's equation~\cite{Sheen1991,Tinkham2004}. Subsequently, we can calculate the magnetic field in the various interfaces surrounding the SQUIDs (where magnetic defects reside) with Biot-Savart's formula, replacing the integral in Eq.~(S14). See details in Supplementary Sec.~\RN{4}B. 

Fits to the model with our numerical extension are shown by the purple and red lines in Fig.~\ref{fig:results}(a,b). We find quantitative agreement with experimental data, including SQUIDs with small wire widths, where the numeric model is consistent with deviations from the linear behavior as observed in experiment. For direct comparison with the analytic approximation, we show the numerical model only including the aluminum-vacuum surface on top of the SQUID (red). As detailed in Fig.~S2, we validate our theoretical model by observing good agreement with the analytical approximation for a small film thickness $b\sim\lambda$ and we confirm that the analytical approximation is inaccurate for our film thickness of $b=\SI{190}{nm}$ and breaks down completely for even higher film thicknesses. Based on our numerical results presented in Fig.~S2, we find that increasing the film thickness $b$ decreases the flux noise amplitude, which is analogous to the effect we observe for increasing the wire width $W$. 

In addition, we perform a fit to the numeric model including defect spins residing in all relevant interfaces surrounding the SQUID, see regions colored in red in the inset schematic in Fig.~\ref{fig:results}(b), i.e. the top and side aluminum-vacuum interfaces, the bottom silicon-aluminum interface, and the silicon-vacuum interfaces beside the SQUID arms, where the magnetic field decays with a power law. Assuming $m=\mu_{\mathrm{B}}$, we obtain $\sigma =\SI{2.6e17}{m^{-2}}$ when considering only the top surface of the SQUID and $\sigma =\SI{6.7e16}{m^{-2}}$ when including all relevant interfaces with equal defect densities. We performed an alternative fit to measured data assuming different defect densities for the aluminum-vacuum, silicon-vacuum, and silicon-aluminum interfaces based on the loss tangents extracted from coplanar waveguide resonators~\cite{Woods2019}, yielding a defect density in the aluminum-vacuum interface of $\SI{1e17}{m^{-2}}$.

The two-dimensional fit to our numerical model including all relevant interfaces is depicted in Fig.~\ref{fig:results}(c), with measured data points shown in Fig.~\ref{fig:results}(d) and relative deviations from the model color-coded. While we measure time-averaged $T_1$ times in our qubits between $\SI{5}{\micro s}$ and $\SI{65}{\micro s}$, with most data points around $\SI{20}{\micro s}$, the extracted values of the noise amplitudes $\sqrt{A_\Phi}$ are in excellent agreement across all measured samples, demonstrating the robustness of our analysis. Dephasing times $T_2$ are limited by $2T_1$ at the sweet spot and are reduced to $\sim\SI{1}{\micro s}$ for the largest frequency detuning from the sweet spot, while $T_1$ times are not limited by flux noise in these samples. Based on the spin echo filter function, our experiment is sensitive to noise frequencies in the range of $\SI{10}{kHz}$ to $\SI{1}{MHz}$.

\begin{table}
\caption{Measured noise amplitudes $\sqrt{A_\Phi}$ for qubits with identical SQUID loops of perimeter $P=\SI{32}{\micro m}$ and wire width $W=\SI{2}{\micro m}$. Fit errors are $\approx\SI{0.1}{\micro\Phi_0}$.}
\label{tab:nice}
{\renewcommand{\arraystretch}{1.2}   
\begin{ruledtabular}
\begin{tabular}{c|ccccccccc}
qubit \#&1&2&3&4&5&6&7&8&9\\
\hline
$\sqrt{A_\Phi}\,(\SI{}{\micro\Phi_0})$&1.46&1.53&1.65&1.70&1.52&1.79&1.70&1.73&1.70\\
\end{tabular}
\end{ruledtabular}
}
\end{table}

Finally, we measure flux noise amplitudes of nine identical qubits with geometry parameters in the optimal limit according to our previous findings. The SQUIDs have small loop perimeters $P=\SI{32}{\micro m}$ and increased wire widths $W=\SI{2}{\micro m}$. These parameters ensure that the three Josephson junctions can be integrated into the SQUID loop without compromising the fabrication quality, although even smaller $P/W$ may be feasible. For the optimized samples, we find consistently low noise amplitudes below $\SI{1.8}{\micro\Phi_0}$, as summarized in Table~\ref{tab:nice}. This verifies the model over a large parameter range and confirms that significant improvements in flux noise levels can be achieved by optimizing SQUID geometry.

The results presented in this Letter are not limited to the specific variant of flux qubit we have used here, but are general to any SQUID used in the framework of superconducting circuits. We substantiate this by measurements of $1/f$ flux noise in capacitively shunted flux qubits where the capacitor is formed by a single floating pad that couples to ground, similar to the `Xmon' layout~\cite{Barends2013}. Both qubit architectures yield consistent flux noise amplitudes for identical SQUID loop geometries. Similarly, the use of ground plane perforations does not result in any trend in noise amplitudes, see Supplementary Sec.~\RN{6}. A summary of the data underlying the results in this Letter is provided in Supplementary Sec.~\RN{7}.

To conclude, we have performed a systematic study of $1/f$ flux noise in more than $50$ capacitively shunted flux qubits with varying SQUID loop geometries and have experimentally demonstrated an approximately linear dependence of the noise power on SQUID perimeter and inverse wire width. Our results are consistent with a model of magnetic two-level defects that reside in the interfaces surrounding the SQUIDs. We have demonstrated quantitative agreement of our data with an extension of the model based on simulating the current distribution in the SQUID loops, resolving the limited applicability and accuracy of the analytic approximation considered previously. This is an important contribution towards solving the long-standing puzzle surrounding the origin of $1/f$ flux noise in conductors. 

The obtained results are expected to be universal for any SQUID-based superconducting circuit. The observed trends -- namely wide wires, small perimeter SQUIDs, and large thickness films being favorable to suppress flux noise -- can therefore serve as a guide to reduce the noise susceptibility of superconducting circuits. In the context of quantum information, this has a direct relevance for improving operational fidelities in both gate-model and quantum annealing approaches to quantum computing.

\section*{Acknowledgments}

The authors are grateful to A. Di Paolo and S. Weber for insightful discussions.

This research was funded in part by the Office of the Director of National Intelligence (ODNI), Intelligence Advanced Research Projects Activity (IARPA), and the Department of Defense (DoD) via MIT Lincoln Laboratory under Air Force Contract No. FA8721-05-C-0002. The views and conclusions contained herein are those of the authors and should not be interpreted as necessarily representing the official policies or endorsements, either expressed or implied, of the ODNI, IARPA, the DoD, or the U.S. Government.

\bibliography{flux_noise}

\begin{thebibliography}{35}
\bibitem [{opt()}]{optcurrent}%
  \BibitemOpen
  \href@noop {} {}\bibinfo {howpublished} {The current distribution across the
  width of the SQUID arms only enters the flux noise amplitude through the
  factor $\int \totd x K(x)^2/(\int \totd x K(x))^2$. Minimizing this
  expression yields that a uniform current distribution $K(x)=\mathrm{const.}$
  minimizes flux noise.}\BibitemShut {Stop}%
\end{thebibliography}
\end{document}


\title{Supplementary material for ``Characterizing and optimizing qubit coherence based on SQUID geometry''}

\def\RLEaffil{Research Laboratory of Electronics, Massachusetts Institute of Technology, Cambridge, MA 02139, USA}
\def\LLaffil{MIT Lincoln Laboratory, Lexington, MA 02421, USA}
\def\Physaffil{Department of Physics, Massachusetts Institute of Technology, Cambridge, MA 02139, USA}
\def\EECSaffil{Department of Electrical Engineering and Computer Science, Massachusetts Institute of Technology, Cambridge, MA 02139, USA}

\author{Jochen Braum\"uller$^{\dagger}$}
\thanks{These two authors contributed equally.\\$^{\dagger}$ jbraum@mit.edu}
\author{Leon Ding}
\thanks{These two authors contributed equally.\\$^{\dagger}$ jbraum@mit.edu}
\author{Antti Veps\"al\"ainen}
\author{Youngkyu Sung}
\author{Morten Kjaergaard}
\affiliation{\RLEaffil}
\author{Tim Menke}
\affiliation{\RLEaffil}
\affiliation{\Physaffil}
\author{Roni Winik}
\affiliation{\RLEaffil}
\author{David Kim}
\author{Bethany M. Niedzielski}
\author{Alexander Melville}
\author{Jonilyn L. Yoder}
\author{Cyrus F. Hirjibehedin}
\affiliation{\LLaffil}
\author{Terry Orlando}
\author{Simon Gustavsson}
\affiliation{\RLEaffil}
\author{William D. Oliver}
\affiliation{\RLEaffil}
\affiliation{\Physaffil}
\affiliation{\LLaffil}
\affiliation{\EECSaffil}

\date{\today}

\newcommand*{\I}{\mathrm{i}}
\newcommand*{\totd}{\text{d}}

\newcommand{\RN}[1]{\textup{\uppercase\expandafter{\romannumeral#1}}}

\maketitle

\section{Sample fabrication}\label{sec:fab}

The samples are fabricated on a silicon substrate by dry etching an MBE grown, $\SI{250}{nm}$ thick aluminum film in an optical lithography process and then diced into $5\times\SI{5}{mm^2}$ chips. Figure~1(b) in the main text shows one of the qubits with its two floating capacitor pads colored in blue. The SQUIDs colored in red in Fig.~1(c) in the main text are fabricated with an electron beam lithography process and a double angle shadow evaporation technique~\cite{Dolan1977} to form the Josephson junctions. Across the entire area of the SQUIDs we evaporate $\SI{40}{nm}$ and $\SI{150}{nm}$ thick aluminum films, separated by an oxide layer created with a controlled in-situ oxidation. In addition to the desired Josephson junctions, this step also creates a large parasitic oxide layer between the aluminum films, which has been shown to host electric two-level systems that lead to qubit decoherence~\cite{Lisenfeld2019}.

\section{Circuit parameters}\label{sec:circuitparams}

A schematic of one qubit-resonator pair coupled to the common transmission line is shown in Fig.~1(e) in the main text. The flux qubit consists of a small Josephson junction that is connected in parallel with two larger Josephson junctions with a relative area ratio of $0.42$. We observe a mean qubit transition frequency of $\omega/2\pi=\SI{4.6}{GHz}$ and a qubit anharmonicity of approximately $\SI{480}{MHz}$ at the qubit sweet spot, located at a flux bias $\Phi$ corresponding to odd integer multiples of half-flux-quantum $\Phi_0/2$, where $\Phi_0 \equiv h/2e$. We explain the observed standard deviation of $\SI{400}{MHz}$ in the sweet spot qubit frequencies in part by geometry dependent variations in the kinetic inductances of the SQUID loops, which are caused by the aforementioned additional parasitic Josephson junctions that form as a result of the shadow evaporation fabrication technique. Additionally, we attribute variations in the qubit transition frequency to a varying junction barrier transparency and small deviations in the junction area. The qubits have a shunt capacitance of $\SI{56}{fF}$ and a critical current density of $\SI{2.4}{\micro A/\micro m^2}$, resulting in $E_{\mathrm{J}}/h=\SI{36}{GHz}$ for the small Josephson junction.

\section{Echo pure dephasing rate due to $1/f$ flux noise}\label{sec:echo}

The qubit is described by the Hamiltonian $\hat H=\hbar\left[\omega_0+\omega(t)\right]\hat\sigma _z/2$, where $\omega_0$ describes a static offset of the qubit frequency and $\omega(t)$ is a stochastic frequency fluctuation induced by flux noise. The solution to the Schr\"odinger equation $\I\hbar\frac{\partial}{\partial t}\Ket\psi=\hat H\Ket\psi$ is
\begin{align}
\Ket{\psi(t)}&=e^{\I\phi(t)}\Ket{\psi(0)}=e^{\I\phi_0+ \I\varphi(t)}\Ket{\psi(0)}\\
\phi(t)&=\omega_0t+\int_0^t\totd t'\omega(t').
\label{eq:phiintc}
\end{align}
We therefore find the stochastic mean of the qubit state to be $\langle\Ket{\psi(t)}\rangle=e^{\I\phi_0}\langle e^{\I\varphi(t)}\rangle\Ket{\psi(0)}$, with the statistical phase accumulation~\cite{Paladino2014}
\begin{equation}
\varphi(t)=\int_0^t\totd t'\omega(t').
\label{eq:varphi}
\end{equation}
By assuming a zero-mean Gaussian distribution~\cite{Makhlin2004,Ithier2005}, the ensemble-averaged phase factor exponential becomes
\begin{equation}
\langle e^{\I\varphi}\rangle=e^{-\langle\varphi^2\rangle/2}.
\label{eq:varphistatmean}
\end{equation}
Within the Gaussian approximation, the statistics of the system are entirely captured by second-order cumulants and we find the phase variance from Eq.~\eqref{eq:varphi}
\begin{equation}
\langle\varphi^2\rangle=\int_0^t\totd t_1\int_0^t\totd t_2 \langle\omega(t_1)\omega(t_2)\rangle.
\end{equation}

In a spin echo experiment, the central $\pi$-pulse effectively inverts the time evolution and we can express the phase accumulation as
\begin{equation}
\tilde\varphi(t)=\left(\int_0 ^{t/2}\totd t'-\int_{t/2} ^t\totd t'\right)\omega(t').
\end{equation}
By using the definition of the noise power spectral density (PSD) $S_\omega(\omega)=\frac 1{2\pi}\int_{-\infty}^\infty \totd\tau e^{-\I\omega\tau}\langle \omega(\tau)\omega(0)\rangle$ with stationary cumulant, the variance becomes
\begin{equation}
\langle\tilde\varphi ^2\rangle=4\int_{-\infty}^\infty \totd\omega S_\omega(\omega) \left(\frac{\sin^2\frac{\omega t}4}{\omega/2}\right)^2 =8\int_0^\infty \totd\omega S_\omega(\omega) \left(\frac{\sin^2\frac{\omega t}4}{\omega/2}\right)^2.
\end{equation}
From Eq.~\eqref{eq:varphistatmean}, we obtain the dephasing component for the spin echo experiment
\begin{equation}
\langle e^{\I\tilde\varphi}\rangle_{\mathrm{E}}=\exp\left(-t^2\int_0^\infty \totd\omega S_\omega(\omega) g_{\mathrm{E}}(\omega,t)\right),
\label{eq:echof0}
\end{equation}
defining the filter function of the spin echo experiment
\begin{equation}
g_{\mathrm{E}}(\omega,t)=\left(\frac{\sin^2\frac{\omega t}4}{\omega t/4}\right)^2.
\end{equation}
For a linear coupling of the noise source to the qubit, we can express the PSD in terms of flux $\Phi$, $S_\omega(\omega)=(\partial\omega/\partial\Phi)^2S_\Phi(\omega)$, and assume a $1/f$ scaling $S_\Phi(\omega)\equiv S(\omega)=A_\Phi/|\omega|$ to obtain
\begin{equation}
\langle e^{\I\tilde\varphi}\rangle_{\mathrm{E}}=\exp\left(-t^2 \left(\frac{\partial\omega}{\partial\Phi}\right)^2 A_\Phi \ln2\right).
\label{eq:dephasingformula}
\end{equation}
The dephasing component in the echo experiment therefore has Gaussian lineshape and we find a pure dephasing rate
\begin{equation}
\Gamma_\phi^{\mathrm{E}}=\sqrt{A_\Phi \ln2} \left|\frac{\partial\omega}{\partial\Phi}\right|.
\label{eq:dephasingVsQubitDeriv}
\end{equation}

In order to compare measured data with our model, we extract the measured flux variance with the inverse Fourier transform of the PSD at $\tau=0$, weighted by the filter function $g_{\mathrm{E}}(\omega,t)$ of the echo experiment for $t>0$,
\begin{equation}
\langle\Phi^2\rangle_{\mathrm{inferred}} =A_\Phi\int_{-\infty}^{\infty}\frac{\totd\omega}{|\omega|}g_{\mathrm{E}}(\omega) =2 A_\Phi\int_{0}^{\infty}\frac{\totd\omega}{\omega}g_{\mathrm{E}}(\omega) =2A_\Phi\ln2.
\end{equation}

\section{Flux noise model based on magnetic defects}\label{sec:model}

We assume non-interacting magnetic two-level system (TLS) defects distributed uniformly across the interfaces surrounding the SQUID, such as the aluminum-vacuum, silicon-vacuum, and silicon-aluminum interfaces. These TLS impurities are modeled as spins with magnetic moment $m$, each coupled to the SQUID through a flux mediated mutual inductance~\cite{Koch2007b, Bialczak2007}. If some test current $I_\text{SQ}$ in the SQUID creates a magnetic field $\vec{B}$ at the location of a TLS, then the flux induced in the SQUID by this TLS is given by $\Phi_\text{SQ} = \vec{B} \cdot \vec{m}_\text{TLS} / I_\text{SQ}$. Assuming a random angular distribution of TLS moments, we calculate the total flux variance as
\begin{equation}
\langle \Phi^2 \rangle = \frac{1}{3} \sigma m^2 \int \totd A \left( \frac{B}{I} \right)^2, 
\end{equation}
where $\sigma$ is the areal density of defects and the surface integral is taken over all considered SQUID interfaces. The factor $1/3$ comes from averaging over the random spin orientations.

The geometries for a toroidal SQUID and a circular SQUID with a rectangular cross-section have been treated by Bialczak \textit{et al.}~\cite{Bialczak2007}. For the SQUID geometry in our experiment -- a rectangular SQUID with rectangular cross-section -- we find
\begin{equation}
\langle \Phi^2 \rangle = \frac{1}{3} m^2\sigma P \int\totd x \left(\frac{B(x)}{I}\right)^2,
\label{eq:flux_var_model}
\end{equation}
where $P = 2X + 2Y + 4W$ is the SQUID perimeter with $X$, $Y$, and $W$ being the SQUID dimensions as shown in Fig.~1 in the main text. The remaining integral, parameterized by $x$, is taken over the lengths of the considered interfaces perpedicular to the extension of the SQUID arms. 

\subsection{Analytic approximation}\label{sec:analytic}

The analytic approximation of the model only considers the top surface of the SQUID as a host for magnetic defects, assuming an effectively two-dimensional film with $W\gg b$. In the integrals involved in this calculation, $x=0$ is treated as the center of the SQUID wire and $x = \pm W/2$ correspond to its two edges. Assuming that the superconducting current flows only at the SQUID surface, the magnetic field at the surface is given by $B(x) = \mu_0 K(x) / 2$ where $K(x)$ is the surface current density. We use a surface current density proportional to $1 / \sqrt{1 - (2x / W)^2}$~\cite{Rhoderick1962} away from the edges, joined by an exponential near the edges at $x = \pm (W/2 - \lambda^2 / 2b)$. Enforcing that the current density has continuous slope, we obtain the following function for the current density along the width of the SQUID
\begin{equation}
K(\bar{x}) = 
\begin{cases} 
K_0 \frac{1}{\sqrt{1 - (2 \bar{x})^2}} & |\bar{x}| \le (1 - \epsilon)/2 \\
K_0 \sqrt{\frac{e}{2 \epsilon}} \exp \left[\left(|\bar{x}| - \frac{1}{2} \right)/ \epsilon \right] & (1 - \epsilon)/2  < |\bar{x}| \le 1/2 
\end{cases}
\label{surface_currents}
\end{equation}
where $\bar{x} \equiv x/W$ and $\epsilon \equiv \lambda^2 / b W$. Since $\epsilon\ll 1$, we keep only leading order terms in $\epsilon$. Evaluating Eq.~\eqref{eq:flux_var_model} with this particular $B(x)$ and using the definition $I=\int\totd x K(x)$, we obtain
\begin{align}
\langle \Phi^2 \rangle &= \frac13 \sigma m^2 P \left(\frac{\mu_0}2\right)^2 \frac{\int \totd x K(x)^2}{(\int \totd x  K(x))^2}\\
&= \frac{\mu_0^2}{3 \pi} m^2 \sigma \frac{P}{W} \left( \frac{\ln(2 b W / \lambda^2)}{2 \pi} + \frac{e - 1}{2 \pi} \right).
\end{align}
The important trends are that the flux variance, and therefore the flux noise power, increases linearly with the average perimeter $P$ of the SQUID loop and decreases roughly inversely with its width $W$.

\subsection{Numerical computation of the current distribution in a superconducting strip of finite thickness}\label{sec:numerics}

\begin{figure} 
\includegraphics{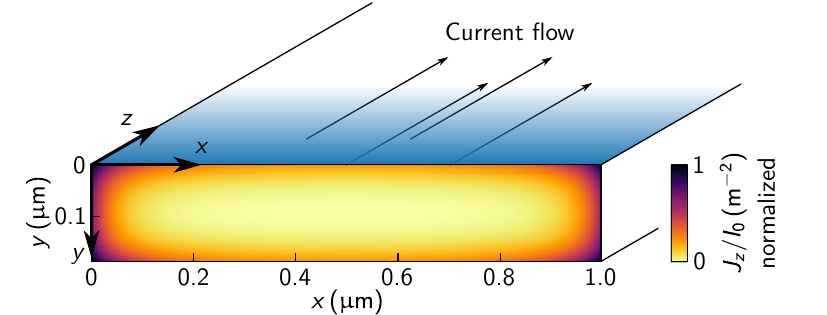}
\caption{Schematic drawing of a section of the SQUID loop in our experiment. Current is flowing along the $z$-direction and we plot the numerically simulated current distribution $J_z/I_0$ at its cross-section, normalized to a reference current $I_0$. Simulation parameters are a standard design with wire width $W=\SI{1}{\micro m}$ and film thickness $b=\SI{190}{nm}$. In order to compute the magnetic field on the surface of the strip (blue) that enters in the flux noise model Eq.~\eqref{eq:flux_var_model}, we use Biot-Savart's law.} 
\label{fig:strip}
\end{figure}

We numerically compute the current distribution in a superconducting strip following the approach presented in Weeks et al.~\cite{Weeks1979} and its extension by Sheen et al.~\cite{Sheen1991}. It is based on the two-fluid model of superconductivity~\cite{Tinkham2004}, where a complex conductivity accounts for both the resistive loss channel at non-zero frequencies (real part) as well as the kinetic energy of the supercurrent (imaginary part). The normal current is described by Ohm's law while the kinetic contribution is added through London's equation. The superconducting penetration depth $\lambda$ enters via the complex conductivity.

We apply the method to calculate the current distribution in a single superconducting strip that is extended in the $z$-direction and discretized into an appropriate number of parallel patches in the $xy$ plane, see Fig.~\ref{fig:strip}. 
We extract the current distribution from the transmission line equation
\begin{equation}
\vec I(\omega)\propto-\I\omega\hat y\vec V,
\end{equation}
where all voltages are set to an identical value (unity). The admittance matrix $\hat y$ is comprised of a resistive part and an inductive part, which in turn depend on the complex conductivity and partial inductances that contain the model geometry~\cite{Weeks1979}. By dividing the currents $I$ penetrating each patch by their cross-sectional area we readily find the volume current density $J(x,y)$ in the $xy$-plane.

Subsequently, we find the magnetic field $\vec B(x)$ on the surface of the strip (blue region in Fig.~\ref{fig:strip}) with Biot-Savart's formula by integrating the current density $J(x,y)$ over the volume of the strip,
\begin{equation}
\vec B(\vec r)=\frac{\mu_0}{4\pi}\iiint _V\totd V\frac{J(x,y)\hat z\times\pvec r '}{|\pvec r '|^3},
\end{equation}
where $\hat z$ denotes the unit vector along $z$ and $\pvec r '$ is the vector from any point in the integration volume to the point where the field is being computed. The norm of this magnetic field, which is mainly directed along $x$ except for some contribution along $y$ close to the edges, enters our flux noise model Eq.~\eqref{eq:flux_var_model}, where it is integrated again along the $x$-dimension. The integrations are numerically approximated by Riemann sums. We chose different numerical discretizations in the successive integrations along $x$ in order to exclude a systematic error and verified our numerical procedure by observing its convergence.

\begin{figure} 
\includegraphics{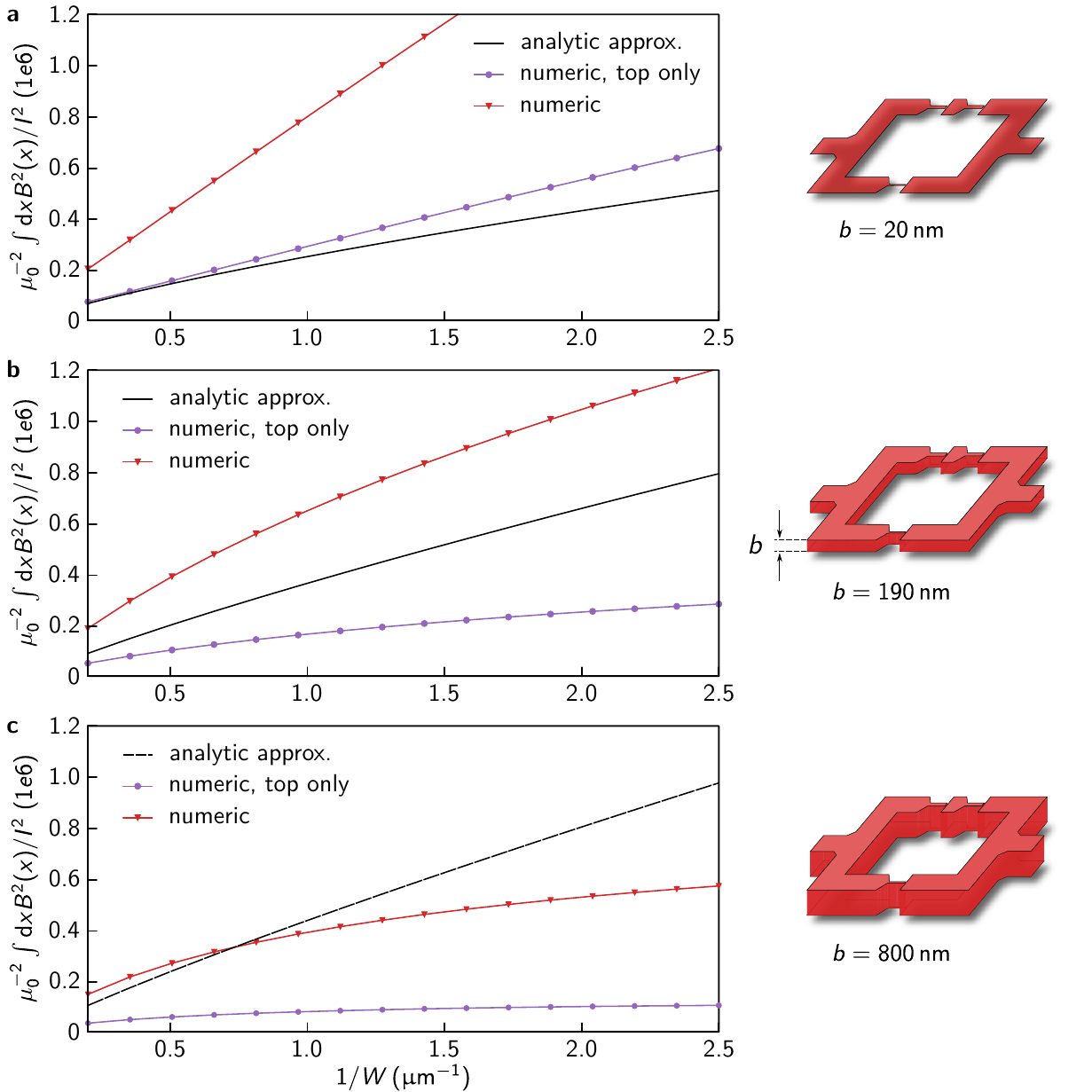}
\caption{Numerical simulation results for various film thicknesses $b$. We plot the numerical value $\mu_0^{-2}\int\totd x B^2(x)/I^2$ which enters the model for magnetic flux noise in Eq.~\eqref{eq:flux_var_model}, replacing the analytic approximation. The purple lines and circles show the results for only considering the top surface of the SQUID and the red lines with triangles show the result for both the top surface and the side faces summed. The analytic approximation is given as a black line. (a) Results for a very thin aluminum thickness $b=\SI{20}{nm}$, where the numerical simulation agrees with the analytic formula for $W\gg b, \lambda$. (b) Results for the thickness $b=\SI{190}{nm}$ as used in our experiment. (c) The analytic expression breaks down entirely for a thick film with $b=\SI{800}{nm}$, while the numerical result indicates a reduced noise sensitivity as compared to smaller wire thicknesses.}
\label{fig:numerics}
\end{figure}

In Fig.~\ref{fig:numerics} we show the results of the numerical simulations for different thickness regimes of the superconducting film forming the SQUID. For each thickness $b$, we plot the numerically obtained factor $\int\totd x B^2(x)/I^2$, a measure for the normalized magnetic field variance, versus the inverse wire width $1/W$. We show simulation results accounting for defect spins located only in the top surface of the SQUID (purple), and we also plot the numerical results taking into account all relevant interfaces (red), which are the top and bottom surfaces of the SQUID arms, their side faces, and the silicon-vacuum interfaces in the vicinity of the SQUID. The numerical simulation of the top surface can be directly compared to the analytic approximation, which yields for the normalized magnetic field variance
\begin{equation}
\int\totd x B^2(x)/I^2=\frac{\mu_0^2}{\pi W} \left(\frac{\ln(2 b W / \lambda^2)}{2 \pi} + \frac{e - 1}{2 \pi} \right),
\end{equation}
see Supplementary Sec.~\ref{sec:analytic}.

We first verify our numerical approach by comparing simulation results for a thin strip with $b=\SI{20}{nm}$ to the result obtained with the approximate formula, see Fig.~\ref{fig:numerics}(a). For large wire widths $W\gg b, \lambda$ in particular, the approximation of the analytic formula is valid and it matches the numerical result for considering defect spins only in the top surface. For thin films, the numerical result recovers the expected linear dependence of $\int\totd x B^2(x)/I^2$ on $1/W$ to first order. Small deviations for smaller $W$ reveal the limitation of the approximate formula~\cite{Rhoderick1962}. For the film thickness $b=\SI{190}{nm}$ used in our experiment (see Fig.~\ref{fig:numerics}(b)), the numerical result and the analytic approximation diverge even for large $W$ since the condition $b\sim\lambda$ is violated. For an even larger thickness of $b=\SI{800}{nm}$, the analytic approximation breaks down completely, see Fig.~\ref{fig:numerics}(c). Remarkably, we find that increasing the film thickness $b$ reduces the noise amplitude considerably, an effect analogous to the reduction in noise with increasing wire width $W$. We want to point out that the contribution from the side faces of the SQUID vanishes in the limit of $b\rightarrow 0$ (Fig.~\ref{fig:numerics}(a)).

\begin{table}
\begin{center}
\caption{Measured data underlying the results presented in Fig.~3 and Table~\RN{1} in the main text. In the sample orientation depicted in Fig.~1(a) in the main text, qubit numbers $1-5$ are located to the left of the transmission line and qubit numbers $6-10$ to the right of the transmission line counted from top to bottom, respectively. For each qubit, we list the SQUID geometry parameters $X$, $Y$, and $W$, the qubit transition frequency $\omega/2\pi$ at the optimal bias point, the average relaxation time $\bar{T_1}$ in the measured region of the spectrum, and the extracted noise amplitudes $\sqrt{A_\Phi}$ left and right of the optimal bias point. Missing values are due to a faulty qubit, electric two-level system modes cutting through the qubit spectrum, or have not been measured.}
\label{tab:data}
\vspace*{1mm}
{\renewcommand{\arraystretch}{1.1}   
\begin{ruledtabular}
\begin{tabular}{ccccccccc}
sample&qubit \#&$X\,(\SI{}{\micro m})$&$Y\,(\SI{}{\micro m})$&$W\,(\SI{}{\micro m})$&$\omega/2\pi\,(\mathrm{GHz})$&$\bar{T_1}\,(\SI{}{\micro s})$&$\sqrt{A_\Phi}\,(\SI{}{\micro\Phi_0})$ (left)&$\sqrt{A_\Phi}\,(\SI{}{\micro\Phi_0})$ (right)\\

\hline
Varied Width & 1 & 9.16 & 8 & 2 & 5.03 & 9 & 2.16 & 2.17 \\
& 2 & 9.16 & 8 & 0.5 & 4.29 & 26 & 2.78 & 2.78 \\
& 3 & 9.16 & 8 & 5 & 5.04 & 22 & 1.58 & 1.59 \\
& 4 & 9.16 & 8 & 0.4 & 4.08 & 38 & 2.81 & 2.87 \\
& 5 & 9.16 & 8 & 1 & 4.8 & 15 & 2.44 & 2.48 \\
& 6 & 9.16 & 8 & 5 & 5.61 & 16 & - & 1.66 \\
& 7 & 9.16 & 8 & 2 & 5.03 & 14 & - & 2.12 \\
& 8 & 9.16 & 8 & 1 & 4.62 & 8 & 2.58 & 2.54 \\
& 9 & 9.16 & 8 & 0.5 & 4.37 & 23 & - & 2.74 \\
& 10 & 9.16 & 8 & 0.4 & - & - & - & - \\ 

\hline
Varied Area & 1 & 12.95 & 11.31 & 1 & 4.73 & 7 & 3.08 & 3.07 \\
& 2 & 6.48 & 5.66 & 1 & 4.85 & 9 & 2.63 & 2.62 \\
& 3 & 25.91 & 22.63 & 1 & 4.61 & 13 & 3.65 & 3.83 \\
& 4 & 4.58 & 4 & 1 & 4.79 & 10 & 1.97 & 2.01 \\
& 5 & 9.16 & 8 & 1 & 4.66 & 13 & 2.67 & 2.69 \\
& 6 & 25.91 & 22.63 & 1 & 5.02 & 10 & 3.86 & 3.88 \\
& 7 & 12.95 & 11.31 & 1 & 4.38 & 17 & 2.75 & 2.77 \\
& 8 & 9.16 & 8 & 1 & 4.95 & 14 & 2.66 & 2.64 \\
& 9 & 6.48 & 5.66 & 1 & 4.51 & 16 & 2.29 & 2.28 \\
& 10 & 4.58 & 4 & 1 & 4.76 & 12 & 2.09 & 2.08 \\

\hline
Varied Aspect Ratio & 1 & 24.21 & 3.03 & 1 & 4.56 & 18 & 3.00 & 3.02 \\
& 2 & 8.56 & 8.56 & 1 & 4.23 & 15 & - & 2.30 \\
& 3 & 34.24 & 2.14 & 1 & 4.72 & 11 & 3.24 & 3.24 \\
& 4 & 4.28 & 17.12 & 1 & 4.32 & 17 & 2.29 & 2.26 \\
& 5 & 17.12 & 4.28 & 1 & 4.26 & 14 & 2.61 & 2.60 \\
& 6 & 34.24 & 2.14 & 1 & 4.57 & 15 & 3.57 & 3.60 \\
& 7 & 24.21 & 3.03 & 1 & 4.48 & 16 & 2.88 & 2.88 \\
& 8 & 17.12 & 4.28 & 1 & 4.57 & 16 & 2.64 & 2.73 \\
& 9 & 8.56 & 8.56 & 1 & 4.10 & 16 & 2.29 & 2.31 \\
& 10 & 4.28 & 17.12 & 1 & 4.07 & 17 & 2.41 & 2.39 \\

\hline
Varied $Y$ & 1 & 18.32 & 45.25 & 1 & - & - & - & - \\
& 2 & 18.32 & 8 & 1 & 4.61 & 17 & 2.55 & 2.56 \\
& 3 & 18.32 & 90.51 & 1 & 4.55 & 9 & 4.62 & 4.57 \\
& 4 & 18.32 & 5.66 & 1 & 4.64 & 16 & 2.89 & 2.82 \\
& 5 & 18.32 & 32 & 1 & 4.77 & 13 & 3.54 & 3.65 \\
& 6 & 18.32 & 90.51 & 1 & 4.66 & 8 & 5.19 & 5.18 \\
& 7 & 18.32 & 45.25 & 1 & 4.43 & 10 & 4.26 & 4.29 \\
& 8 & 18.32 & 32 & 1 & 4.21 & 16 & 3.82 & 3.79 \\
& 9 & 18.32 & 8 & 1 & 4.63 & 15 & 2.90 & 2.94 \\
& 10 & 18.32 & 5.66 & 1 & 4.57 & 16 & 3.00 & 3.01 \\

\hline
Identical & 1 & 6.41 & 5.6 & 2 & 5.15 & 17 & 1.46 & 1.45 \\
& 2 & 6.41 & 5.6 & 2 & 5.07 & 17 & 1.55 & 1.50 \\
& 3 & 6.41 & 5.6 & 2 & 4.92 & 17 & 1.67 & 1.63 \\
& 4 & 6.41 & 5.6 & 2 & 4.57 & 18 & 1.69 & 1.70 \\
& 5 & 6.41 & 5.6 & 2 & 4.80 & 17 & 1.49 & 1.54 \\
& 6 & 6.41 & 5.6 & 2 & 3.00 & 45 & 1.93 & 1.83 \\
& 7 & 6.41 & 5.6 & 2 & 4.83 & 18 & 1.79 & 1.79 \\
& 8 & 6.41 & 5.6 & 2 & 4.84 & 15 & 1.69 & 1.70 \\
& 9 & 6.41 & 5.6 & 2 & 4.92 & 14 & 1.72 & 1.74 \\
& 10 & 6.41 & 5.6 & 2 & 4.87 & 17 & 1.71 & 1.68 \\

\end{tabular}
\end{ruledtabular}
}
\end{center}
\end{table}

\begin{table}
\begin{center}
\caption{Measured data of six qubits on one sample with either a floating or grounded shunt capacitor. All other parameters were otherwise kept identical. The data show no trend in the noise amplitudes extracted for the different layouts, indicating that the noise amplitude values are not dependent on qubit architecture but only on SQUID geometry. For each measured qubit, we list the SQUID geometry parameters $X$, $Y$, and $W$, the qubit transition frequency $\omega/2\pi$ at the optimal bias point, the average relaxation time $\bar{T_1}$ in the measured region of the spectrum, and the extracted noise amplitudes $\sqrt{A_\Phi}$ left and right of the optimal bias point. Missing values have not been measured.}
\label{tab:datamixed}
\vspace*{1mm}
{\renewcommand{\arraystretch}{1.1}   
\begin{ruledtabular}
\begin{tabular}{ccccccccc}
sample &$X\,(\SI{}{\micro m})$&$Y\,(\SI{}{\micro m})$&$W\,(\SI{}{\micro m})$&$\omega/2\pi\,(\mathrm{GHz})$&$\bar{T_1}\,(\SI{}{\micro s})$&$\sqrt{A_\Phi}\,(\SI{}{\micro\Phi_0})$ (left)&$\sqrt{A_\Phi}\,(\SI{}{\micro\Phi_0})$ (right)&capacitor shape\\

\hline
Varied Pads & 9.16 & 8 & 1 & 4.62 & 15 & 2.31 & - & floating \\
& 9.16 & 8 & 1 & 4.54 & 12 & 2.39 & - & floating \\
& 9.16 & 8 & 1 & 5.16 & 5 & 2.07 & - & floating \\
& 9.16 & 8 & 1 & 4.04 & 17 & - & 2.18 & grounded \\
& 9.16 & 8 & 1 & 4.67 & 14 & 2.31 & 2.33 & grounded \\
& 9.16 & 8 & 1 & 4.45 & 17 & 2.25 & 2.31 & grounded \\

\end{tabular}
\end{ruledtabular}
}
\end{center}
\end{table}

\section{Extended data set}

\begin{figure} 
\includegraphics{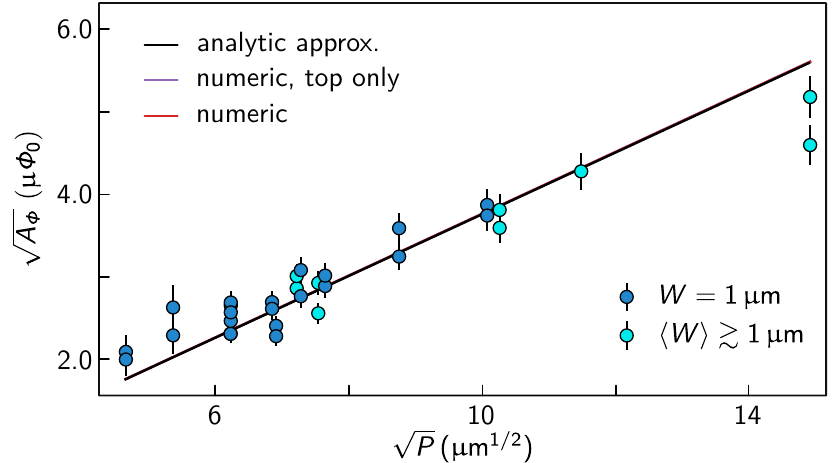}
\caption{Extended data set of measured flux noise amplitudes $\sqrt{A_\Phi}$ in dependence of SQUID perimeter $P$ for a nominally constant wire width $W=\SI{1}{\micro m}$. Dark blue data points are reproduced from Fig.~3(a) in the main text. The inclusion of the light blue points form an extended data set consisting of nine additional SQUIDs from a chip with smaller aspect ratios ($X/Y$). As a result of the shadow angle evaporation technique, the effective width for these data points is increased, yielding a  reduction of flux noise.}
\label{fig:results_extended}
\end{figure}

As discussed previously, the double angle shadow evaporation technique creates an increased width along the vertical arms of the SQUIDs by $\sim\SI{350}{nm}$. In SQUIDs with aspect ratios $X\ll Y$ in particular, this effect results in an average effective width significantly different from the nominal value of $\SI{1}{\micro m}$. In Fig.~3(a) in the main text, we therefore omitted data points for SQUIDs containing these smallest aspect ratios for clarity. Here in Fig.~\ref{fig:results_extended}, we show our complete data set, where data points in light blue correspond to SQUIDs with constant $X$ but the vertical dimension $Y$ of the SQUID variably increased. Apart from the slight decrease of the measured $\sqrt{A_\Phi}$ due to the increased width, the expected linear scaling of the theory is well reproduced in this extended parameter range.

\section{Ground plane perforations}\label{sec:fluxtraps}

We have used chip designs for this noise study with $\SI{3}{\micro m} \times \SI{3}{\micro m}$ sized ground plane perforations acting as flux traps, distributed at a pitch of $\SI{13}{\micro m}$ (center-to-center distance). We did not observe a dependence of the flux noise amplitudes on the presence or absence of ground plane perforations. During the course of this experiment, we noticed flux instabilities on a timescale of seconds and nonlinear flux tuning for bias currents below $\sim\SI{1}{mA}$, but stable conditions for larger bias currents. In order to operate in the stable regime, we chose optimal bias points corresponding to larger bias currents to perform the flux noise study. We speculate that this behavior is caused by a reconfiguration of weakly pinned flux vortices.

\section{Complete data set}\label{sec:data}

Measured data underlying our results is listed in Table~\ref{tab:data} and Table~\ref{tab:datamixed}.

\bibliography{flux_noise}